\documentclass[a4paper,10pt,twoside]{cpc-hepnp}
\usepackage{epsfig}
\usepackage{url}
\usepackage{multicol}
\usepackage{graphicx}
\usepackage{booktabs}
\usepackage{amssymb,bm,mathrsfs,bbm,amscd}
\usepackage[tbtags]{amsmath}
\usepackage{lastpage}

\begin{document}
\newcommand{\DAF}{DA$\rm{\Phi}$NE}
\newcommand{\g}{\gamma}
\newcommand{\ee}{e^+e^-}
\def\pbi{pb$^{-1}$}
\def\fbi{fb$^{-1}$}

\fancyhead[co]{\footnotesize G.Venanzoni: Status of KLOE-2}


\title{Status of  KLOE-2}

\author{%
      G.~Venanzoni$^{1)}$\email{graziano.venanzoni@lnf.infn.it}\\
      for the KLOE-2 Collaboration~\cite{kloe2}}
\maketitle

\address{%
Laboratori Nazionali di Frascati dell'INFN, I-00044, Frascati, Italy\\
}

\begin{abstract}
In a few months 
the KLOE-2 detector is expected to start data taking at the upgraded
\DAF\ $\phi$-factory of INFN Laboratori Nazionali di Frascati.
It aims to collect 25 fb$^{-1}$ at the $\phi(1020)$ peak, 
and about 5 fb$^{-1}$ in the energy region between 1 and 2.5 GeV. 
We review the status and physics program of the project.
\end{abstract}


\begin{multicols}{2}

\section{Introduction}
\par\noindent
From 2000 to 2006 the KLOE experiment has collected 2.5 fb$^{-1}$ of
data at the $\phi(1020)$ peak plus additional 250 pb$^{-1}$ off-peak at the 
\DAF\ $\phi$-factory of INFN Laboratori Nazionali di Frascati.
Many important results have been obtained,
particularly in the kaon sector, light meson spectroscopy and 
 on the precise measurement of the hadronic cross section below 1 GeV.
During 2008 a new interaction scheme of DA$\Phi$NE has
been successfully tested, allowing to reach a peak luminosity of about 5$\times
10^{32}$ cm$^{-2}$ s$^{-1}$, a factor of 3 larger than previously
obtained. 
Following this achievement, new data taking with an upgraded
detector will start in 2010.
To extend the precise measurement of the hadronic cross sections 
above 1 GeV  a program for running \DAF\ at energies up to 2.5
GeV has also been suggested.
The improved KLOE detector is perfectly suited for taking data at 
energies different from the $\phi$ peak, while \DAF\
will need some upgrade~\cite{dafne2}.

We will refer to the entire plan of run discussed above as
the KLOE-2 project.

\section{The KLOE-2 detector}
KLOE is a general purpose detector for $e^+ e^-$ physics, consisting mainly
of a large cylindrical (helium based) drift chamber, with an internal radius of 25 cm and an external one of 2 m, surrounded
by a lead-scintillating fiber electromagnetic calorimeter embedded in a superconducting magnet ($B=0.52$ T).
The detector design was optimized for $CP$ studies in the neutral kaon system, for kaons produced in the decay of $\phi$ almost at rest.
The upgrade of the KLOE detector consisting in the installation of 
an electron tagger for $\gamma\gamma$ physics has been completely funded.
Two different tagging detectors will be installed: the Low Energy Tagger, (low
energy refers to the $e^{\pm}$ energy) LET, made of two crystal
calorimeters placed close to the DA$\Phi$NE Interaction Point (IP) in
symmetrical positions, and the High Energy Tagger, HET, made of two position
sensitive detectors placed (symmetrically) far from the IP, after the first
bending dipoles of DA$\Phi$NE. 
\begin{center}
\includegraphics[width=8cm,height=5cm]{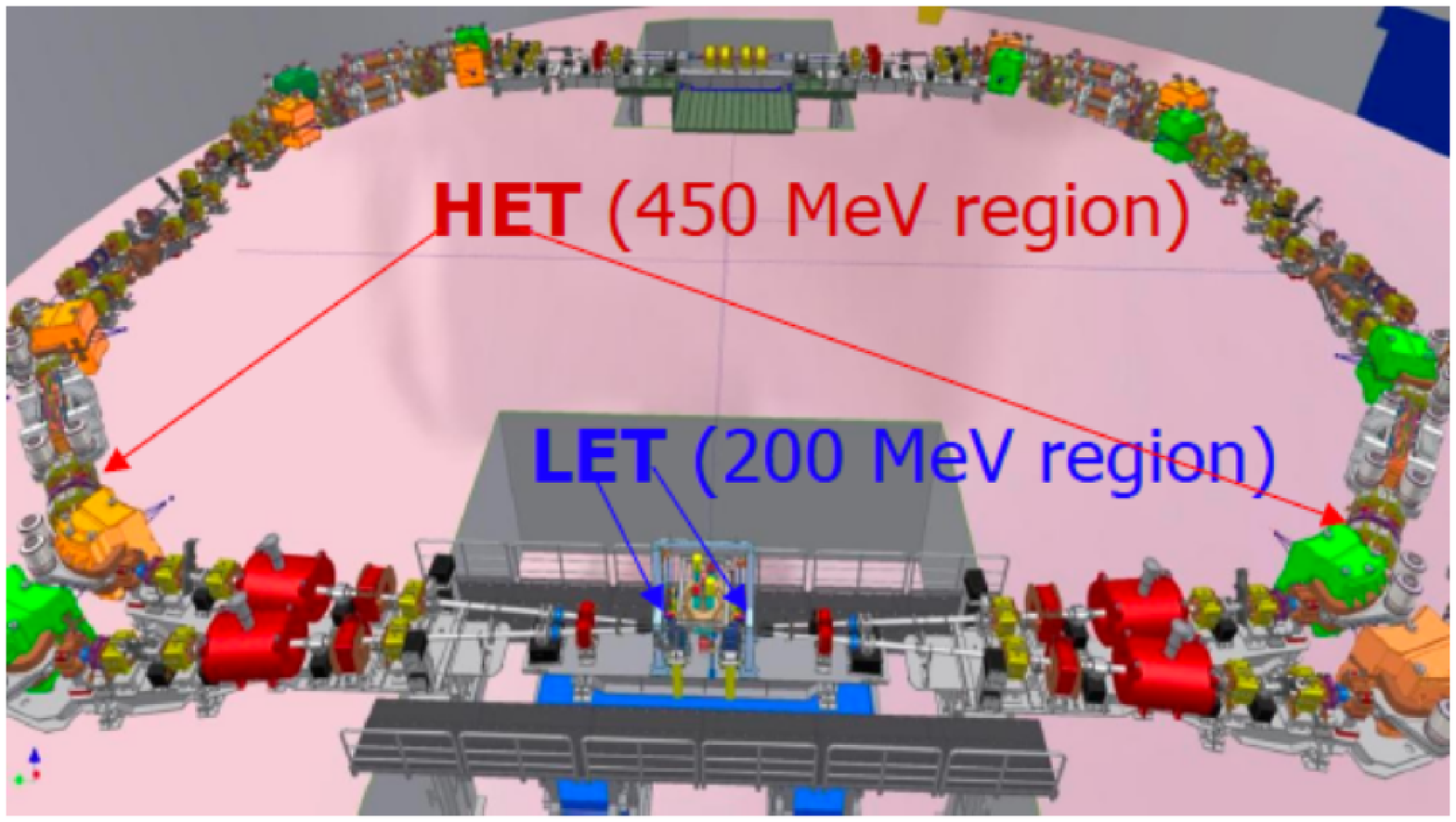}
\end{center}
\figcaption{Layout of \DAF\ with the positions of  LET and HET detectors for
 $\gamma\gamma$ physics. The average $e^{\pm}$ energy for each detector is shown.}
\label{fig:gg}
Figure~\ref{fig:gg} shows the \DAF\ layout with the positions of LET and HET detectors for $\gamma\gamma$ physics. 

A further detector upgrade has been proposed for a second phase of the
data taking, with the insertion of a light internal tracker between the
beam pipe and the drift chamber, and crystal calorimeters to cover the
polar angle regions down to 8$^{\circ}$.  This upgrade will be very 
important for the kaon interferometry and multihadronic cross section measurements (see below).
 It has been partially funded and the installation is envisaged for late 2011.

\section{The KLOE-2 physics program}
The KLOE-2 program covers a wide variety of physics topics: 
tests of CKM Unitarity and Lepton Universality with kaons, tests of
discrete symmetries and of Quantum Mechanics with entangled kaon states,
rare kaon decays, light meson spectroscopy (scalar and pseudoscalar
mesons), measurement of the hadronic cross section from the $2m_{\pi}$ threshold to 2.5 GeV, search for possible
Dark Matter signals at low energy. 
In the following we will summarise some of the above topics. For a detailed discussion see~\cite{kloe2paper}.

\subsection{Kaon interferometry and CPT tests}
$CPT$ invariance is a fundamental theorem in quantum field theory.
In several quantum gravity (QG) models, however, $CPT$ 
can be violated
via some  mechanism which can also 
violate standard Quantum Mechanics (QM). 
In this respect the entagled neutral kaon 
pairs produced at \DAF\  play a unique role
in precision tests of the $CPT$ symmetry~\cite{bemp}.
As an example of this incredible precision reachable with neutral kaons,
let us consider the model by Ellis, Hagelin, Nanopoulos and Srednicki (EHNS)
 which introduces three $CPT$ and QM-violating
real parameters $\alpha$ $\beta$ and $\gamma$~\cite{ehns}.  
On phenomenological grounds, they are expected to 
be \break
 $O(m^{2}_{K}/M_{Pl})\sim2\times$10$^{-20}$GeV at most, being
$M_{Pl}\sim10^{19}$ GeV, the Planck mass. 
Interestingly enough, this model gives
 rise to observable effects in the behaviour
 of entagled neutral meson systems, as shown also in \cite{phu},  
 that can be experimentally tested.
With 2.5 $fb^{-1}$ 
KLOE has published competitive results on these issues~\cite{DiDomenico:2009zz}.
The analysis makes use of correlated $K^{0}_{L}-K^{0}_{S}$ pairs,  
by measuring the relative distance of their decay point into two charged pions. 
The decay region most sensitive  to the EHNS parameters is the one close to
the IP. 
 
\begin{center}
\includegraphics[width=9.5cm,height=12cm]{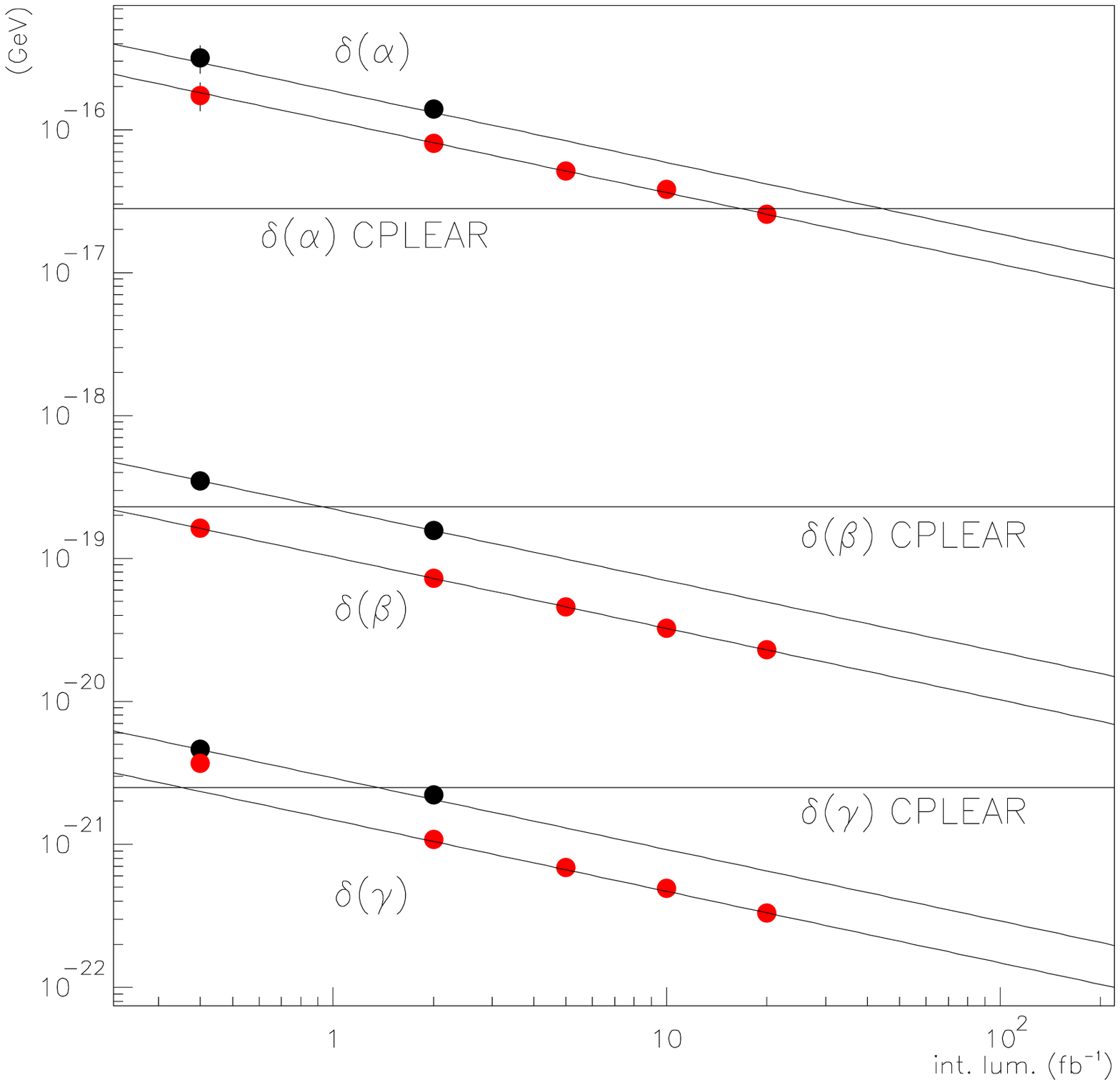}
\end{center}
\vspace{-2.cm}
\figcaption{Limits on the CPT violating parameters 
$\alpha$, $\beta$, and $\gamma$ obtainable by KLOE-2 as a function of the integrated
luminosity. See the text for details.}
\label{fig:qmp}

Figure~\ref{fig:qmp} shows the potential limits that can be obtained by KLOE on 
$\alpha$, $\beta$, and $\gamma$ as a function of the integrated luminosity, 
both with and without the insertion of an inner tracker
with vertex resolution of 0.25~$\tau_{S}$ (to be compared with the present
KLOE vertex 
resolution, 0.9~$\tau_{S}$). In the figure are also given the  
results from CPLEAR~\cite{cplearqm}.
Without giving too many details, it is clear that with a 
reasonable  integrated luminosity, KLOE-2 can set the best limits on these 
parameter.

\subsection{Other tests of discrete symmetries}
CPT symmetry can also be tested using neutral kaon semileptonic decays. 
Actually, the charge asymmetries of the short- and long-lived $K^{0}$ mesons
for these reactions must be equal under CPT  conservation. 
The $K_{L}$ asymmetry is known today with a precision of
 order 10$^{-5}$~\cite{Amsler:2008zzb}, 
while the $K_{S}$ one, measured 
by KLOE using $\sim$ 400 \pbi\ of data, is known at the per cent level
~\cite{ref:kkss}. 
KLOE-2 aims at reaching a 10$^{-3}$ precision.

$K_{S}$ decays into three neutral pions are purely CP violating. 
the branching ratio for this decay is expected to be $\sim$10$^{-9}$. 
KLOE has set the best limit to date on this branching ratio: 
1.2$\times$10$^{-7}$ at 90$\%$ C.L~\cite{ref:ks30}.
KLOE-2 can improve a lot on that, and might hope to observe the signal for the 
first time. Improvements can arrive not only on the statistics side; in fact, 
usage of the forward calorimeters can improve the rejection of the 
$K_{S} \rightarrow 2\pi^{0}$ events, the most relevant systematic 
limitation for the measurement. 

Unconventional forms of CP violation can also be tested using $\eta$ meson 
decays to $\pi^{+}\pi^{-}e^{+}e^{-}$. Actually, thanks to the $\phi$
radiative decay to $\eta\gamma$, \DAF\ can also be considered a 
clean source of well tagged $\eta$ mesons. KLOE has performed 
the measurement of the branching ratio $\eta\to\pi^{+}\pi^{-}e^{+}e^{-}$ with 
4\% accuracy and has obtained the first measurement of the CP-odd $\pi\pi–ee$ decay plane angular asymmetry~\cite{ref:asy}.
About 10$^{9}$ $\eta$'s
will be collected by KLOE-2 in a few years of data taking. With this sample
an asymmetry as small as 10$^{-3}$ can be measured.  

KLOE has already set the best limits on the decay rates of 
$\eta \rightarrow \pi^{+}\pi^{-}$~\cite{ref:kepp} and
$\eta \rightarrow \gamma\gamma\gamma$~\cite{ref:keee},
 two processes which are 
forbidden by invariance under P and C transformations. 
Using the full KLOE-2 statistics one 
can improve the above results by about two orders of magnitude; these will
be the most precise tests ever done of P and C conservation in
strong and electromagnetic interactions.  

\subsection{Light scalar spectroscopy at $\phi(1020)$ peak}
It is still controversial whether light scalars are $q\bar q$
mesons, $qq\bar q\bar q$ states, or $K \bar K$ molecules.
KLOE exploited the radiative decays $\phi\to PP\gamma$ to study $f_0(980)$ and
$a_0(980)$, and to look for a signal of the $\sigma(600)$, and extracted
the parameters of the scalar resonances from the two pseudoscalar invarant
mass distributions~\cite{gauzzi,achasov}. 
Substantial improvements from KLOE-2 are expected for
$\phi\to(f_0/a_0)\gamma\to K^0\bar K^0\gamma$: 
the KLOE upper limit, ${\rm Br}(\phi\to K^0\bar K^0\gamma) < 1.9\times
10^{-8}$\cite{phikk}, can be lowered, with the KLOE-2 statistics, down to
$1\times 10^{-8}$.
This limit can be further reduced to $0.5\times 10^{-8}$ with the insertion of the inner tracker.
This value is in the range of the theoretical predictions for the
branching ratio, then
the first observation of this decay is possible at KLOE-2.\\
In $\eta^{\prime}\to\eta\pi\pi$ decays the $\pi\pi$ system has the same
quantum numbers of a scalar meson. 
Moreover, the available kinetic energy of the $\pi\pi$ system
is in the range (0, 137) MeV, suppressing high angular momentum
contributions, and the exchange of vector mesons is forbidden by G-parity 
conservation. 
For these reasons only scalar mesons can participate in the scattering
amplitude. 
The decay can be mediated by the scalar ($\sigma$,$a_0$ and $f_0$)
exchange and by a direct contact term due to the chiral anomaly
\cite{fariborz}. The scalar contribution can be determined from a fit to
the Dalitz plot. 
A Monte Carlo simulation of the $\eta^{\prime}\to\eta\pi^+\pi^-$ process 
shows that KLOE-2 has a good sensitivity to $\sigma(600)$.

\subsection{$\gamma\gamma$ physics}
The term ``$\gamma\gamma$ physics'' (or ``two-photon physics'') stands 
for the study of the reaction:
$$
\ee\,\to\,\ee \,\g^*\g^*\,\,\to\,\ee \,+\, X
$$
where $X$ is some arbitrary final state allowed by conservations laws. 

The number of $e^+e^-\rightarrow e^+e^-X$ events per unit of invariant mass $W_{\gamma\gamma}$, as a function 
of $W_{\gamma\gamma}$ itself, is:
\begin{eqnarray*}
N({\rm evts/MeV})& = & L_{\rm int}({\rm nb^{-1}}) \times \\
\frac{{\rm d} F (W_{\gamma \gamma},\sqrt{s})}{{\rm d} W_{\gamma\gamma}}
({\rm MeV^{-1}}) 
& \times & \sigma ( \gamma \gamma \rightarrow X) ({\rm nb})
\end{eqnarray*}
where $L_{\rm int}$ is the $e^+e^-$ integrated luminosity and 
${\rm d} F(W_{\gamma\gamma},\sqrt{s})/{\rm d} W_{\gamma\gamma}$ is the effective 
$\gamma\gamma$ luminosity per unit energy. The product ${\rm d} F/{\rm d} W \times L_{\rm int}$ 
is reported in Fig.~\ref{gg} ({\it Left}) for two \DAF\ center-of-mass (c.m.)
 energies.

\subsubsection{The process $\gamma\gamma \rightarrow \pi^o\pi^o$: the $\sigma$ case }
$\gamma\gamma$-physics provides a complementary
 view at the light scalar mesons and, in 
particular, is a powerful tool to search for the
 $\sigma$~\cite{achasov}. $e^+e^-\rightarrow e^+e^-X$ events 
with $X = \pi\pi$, $\eta\pi$ and possibly $K\bar{K}$, 
allow to study directly the $I = 0$ 
and $I = 1$ scalar amplitudes down to their thresholds. 
In $\gamma\gamma \rightarrow \pi^0\pi^0$ 
events with two-photon invariant masses $W_{\gamma\gamma}$ below 1 GeV, 
the $\pi^0\pi^0$ pair is mostly in the S-wave, resulting in $J^{PC} = 0^{++}$
 quantum numbers, with a negligible contamination 
from other hadronic processes.
 The presence of a pole in this amplitude around 500 MeV 
\cite{Caprini} would be a clean and new signal of the $\sigma$.  
 
\begin{figure}[ht]
\centering
\mbox{\epsfig{file=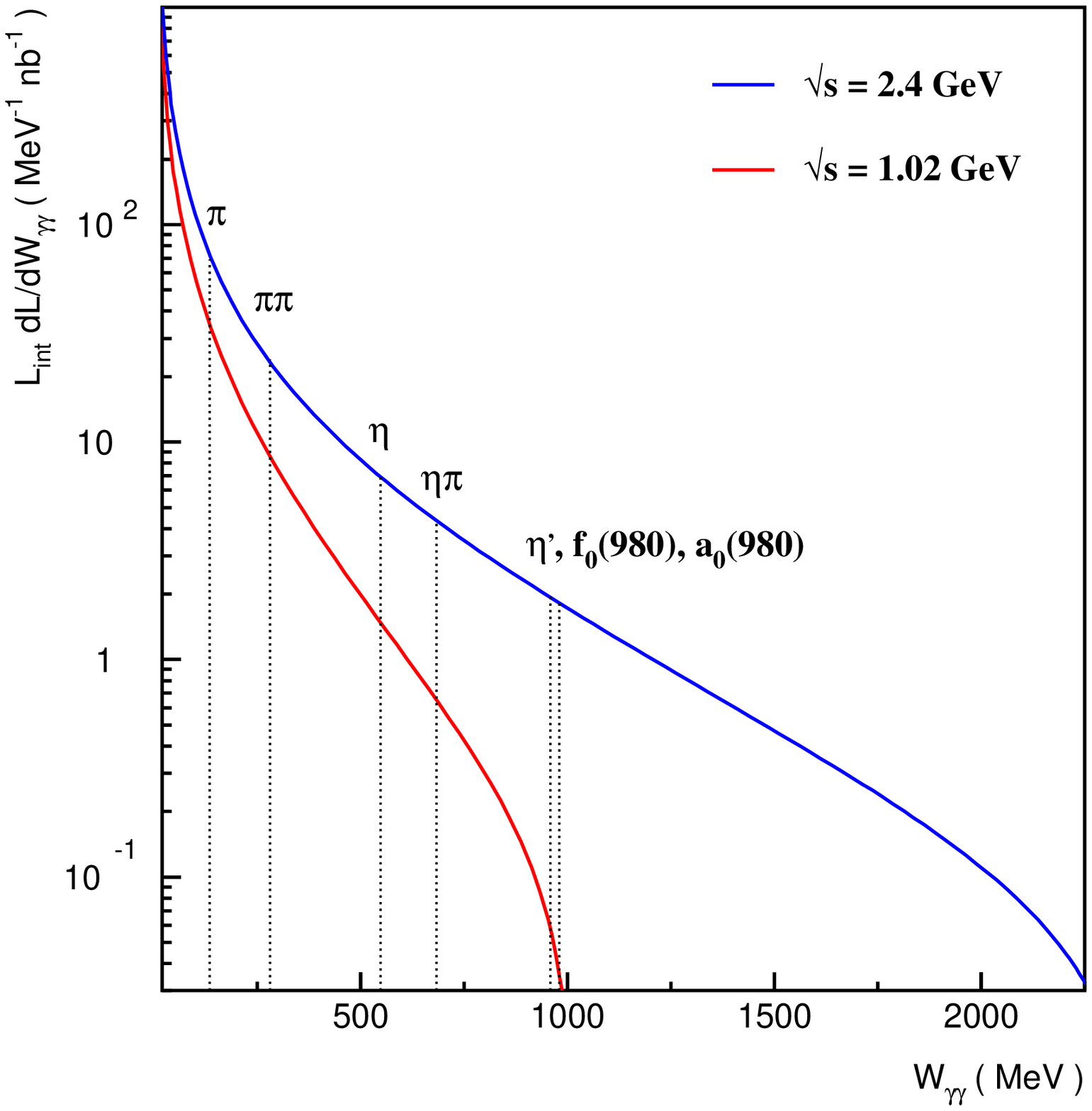,width=6cm,height=6cm}
\epsfig{file=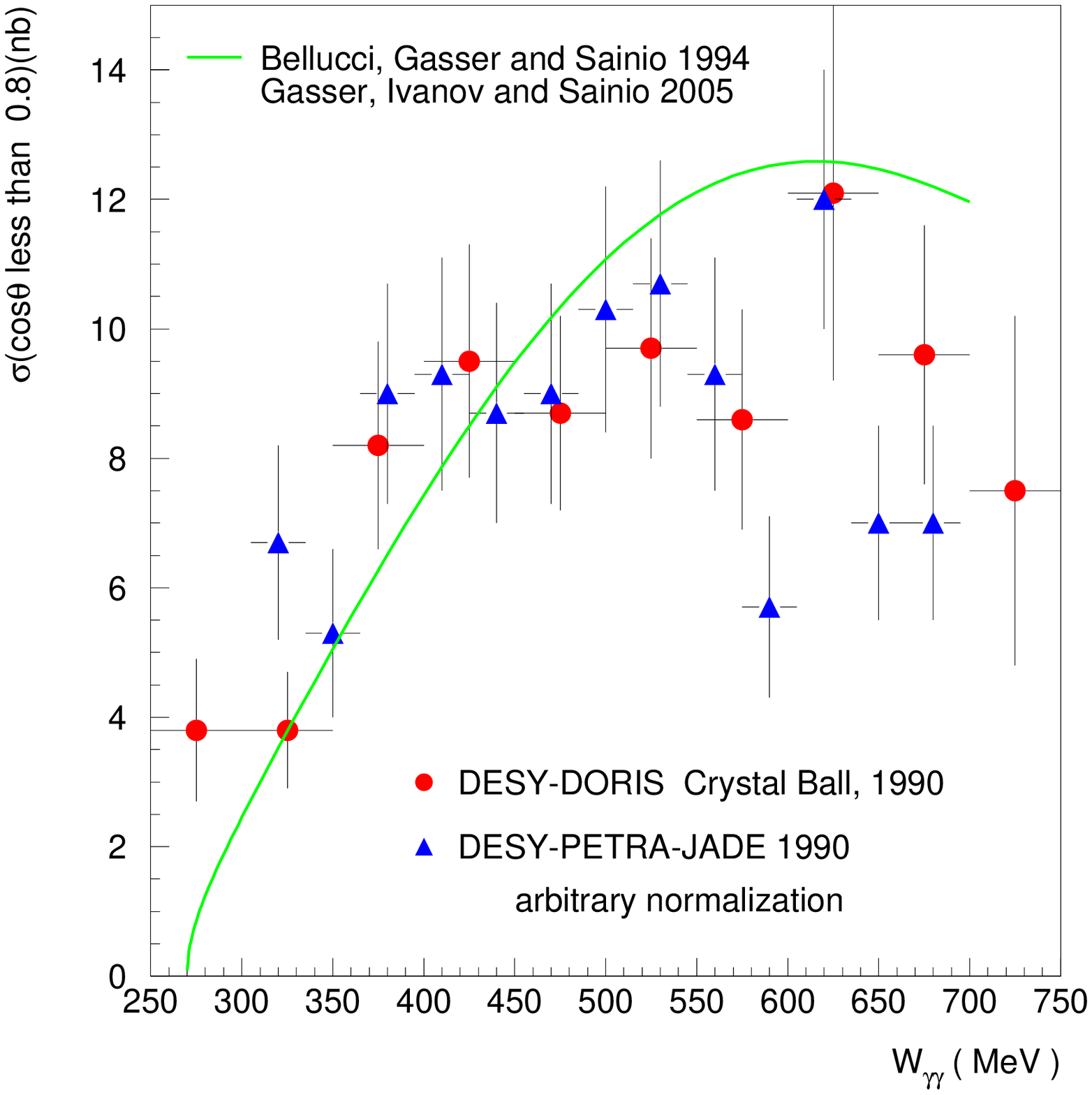,width=6cm,height=6cm}}
\caption{\small{{\it Left}: Effective $\gamma\gamma$ luminosity as a function of
 $W_{\gamma\gamma}$ corresponding to an integrated luminosity 
 of  1 fb$^{-1}$ at $\sqrt{s} = m_{\phi}$ (red curve) and at 
$\sqrt{s}$=2.4 GeV (blue curve). Vertical lines represent from left 
to right: $\pi$-threshold, $\pi \pi$-threshold, $\eta$, $\eta \pi$-threshold,
$\eta'$, $f_0$, $a_0$. {\it Right}: Collection of low energy $\gamma\gamma\rightarrow\pi^{0} \pi^{0}$ 
                cross section data compared with a theoretical evaluation  		    based on $\chi$PT \cite{Bellucci}. The JADE data are normalised 		    to the same average cross section as the Crystal Ball data. 
}}
\label{gg}
\end{figure}

Unfortunately, the only available  experimental information on this 
channel in the region of interest is relatively poor
and does not allow to draw any conclusion about the agreement with either the 
$\chi$PT
calculations or on the  existence of the broad (250-500 MeV) 
$\sigma$ resonance  (see Fig.~\ref{gg} ({\it Right})). 
KLOE is finalizing a new measurement of  $\gamma\gamma \rightarrow \pi^0\pi^0$ in this region using 250 pb$^{-1}$ of data taken at 1 GeV~\cite{nguyen}.
%
%

\end{multicols}
\ruleup
\ruledown
\begin{multicols}{2}

\subsubsection{Measurement of the $\gamma\gamma$ widths of $f_0(980)$ and $a_0(980)$} 
Extending the measurement of $\gamma\gamma \rightarrow \pi\pi$ and 
$\gamma\gamma \rightarrow \eta\pi$ up to $W_{\gamma\gamma}\,\sim\,1$ GeV, the two-photon 
width of $f_0(980)$ and $a_0(980)$ can also be measured. This measurement is possible by 
running at the maximum attainable c.m. energy, in order to maximise
the effective $\gamma\gamma$ luminosity in the GeV region 
(see Fig.~\ref{gg}, {\it Left}).
In both 
cases a peak in the $W_{\gamma\gamma}$ dependence of the 
$\gamma\gamma \rightarrow \pi\pi (\eta\pi)$ cross section around the meson mass allows to 
extract the $\gamma\gamma$-width.

\subsubsection{The two-photon widths of the pseudoscalar mesons}
The
situation wtih the decay constants of  $\eta$ and $\eta'$
is far from being satisfactory and calls for more precise measurements 
of the two-photon width of these mesons~\cite{kloe2paper}. Even the
 $\pi^0$ two-photon width is poorly known (relative
uncertainty of $\sim 8\%$) and its determination can be
improved at \DAF. Given the small value 
of these widths, the only way to measure them is the
 meson formation in $\g\g$ reactions. 
In Table~\ref{ggtab} we report the estimates for the total production rate of 
a pseudoscalar meson (PS) in the process $e^+e^-\rightarrow e^+e^-PS$ 
for two \DAF\ c.m. energies~\cite{kloe2paper}.
\begin{center}
  \tabcaption{\label{ggtab}$e^+e^-\rightarrow e^+e^-PS$ total rate for an integrated
    luminosity of 1 fb$^{-1}$ at two different c.m. energies. No
    tag efficiency is included in the rate calculation.}
\footnotesize
  \begin{tabular*}{80mm}{c@{\extracolsep{\fill}}cccc}
    \toprule
    $\sqrt{s}$ (GeV) & $\pi^0$ & $\eta$ & $\eta'$ \\
    \hline
    1.02 & 4.1$\times 10^5$ & 1.2$\times 10^5$ & 1.9$\times 10^4$ \\
    2.4 & 7.3$\times 10^5$ & 3.7$\times 10^5$ & 3.6$\times 10^5$ \\
    \hline
\bottomrule
  \end{tabular*}
\end{center}

\subsubsection{Meson transition form factors}\label{sec:ffact} 
The process $e^+e^- \to e^+e^- + PS$ with one of the final 
leptons scattered at large angle gives access to the process 
$\g\g^* \to PS$, i.e. with one off-shell photon, and it allows to extract
information on the pseudoscalar meson transition form factor (TFF) $F_{P\g\g^*}(Q^2)$.

By detecting both leptons at large angles the doubly off-shell TFF 
$F_{P\g^*\g^*}(Q^2_1,Q^2_2)$ can be accessed.
A direct and accurate determination of these quantities either with one or both leptons scattered at large angle would be
extremely important in the region below 1-2 GeV, 
where few data are available.
It would also help to get less model-dependent estimations of 
the hadronic  light-by-light contribution 
to $(g-2)_{\mu}$~\cite{nyffeler}.
%

\subsection{Measurement of the hadronic cross sections  in the energy region below 2.5 GeV}
In the last years the improved precision reached in the
 measurement of $e^+e^-$ annihilation cross sections
in the energy range below a few GeV has led to a substantial reduction 
on the uncertainty of the hadronic contribution to the effective fine-structure constant at the scale $M_Z$, $\Delta\alpha^{(5)}_{\rm{had}}(M_Z^2)$, and 
to the anomalous magnetic moment of the muon $a^{\rm{HLO}}_{\mu}$~\cite{davier,teubner}.
However, while below 1 GeV the error on the two-pion channel
which dominates the cross section in this energy range 
is below 1\%, the region
between 1 and 2 GeV is still poorly known, with a fractional
accuracy of $\sim10\%$. Since this region contributes about 40\%~\cite{Jegerlehner:2008rs} 
to the total error of $\Delta\alpha^{(5)}_{\rm{had}}(M_Z^2)$ (and up to $\sim 70\%$ by using the
Adler function as proposed in~\cite{Jegerlehner:2008rs}),
 and about 55\%~\cite{Jegerlehner:20081} to the error of $a^{\rm{HLO}}_{\mu}$, it is evident how desirable an improvement on this region is~\cite{kloe2paper}.

KLOE-2 can play a major role by measuring in this energy region the hadronic cross section at 1-2\% level.
With a specific luminosity of $10^{32}$cm$^{-2}$s$^{-1}$, 
DA$\mathrm{\Phi}$NE upgraded in energy 
can perform a scan in the region from 1 to 2.5 GeV,
collecting an integrated luminosity of 20 pb$^{-1}$ per point (corresponding
to a few days of data taking). By assuming an energy step of 25 MeV, the whole
region would be scanned in one year of data taking.
\begin{center}
\includegraphics[width=8cm,height=8cm]{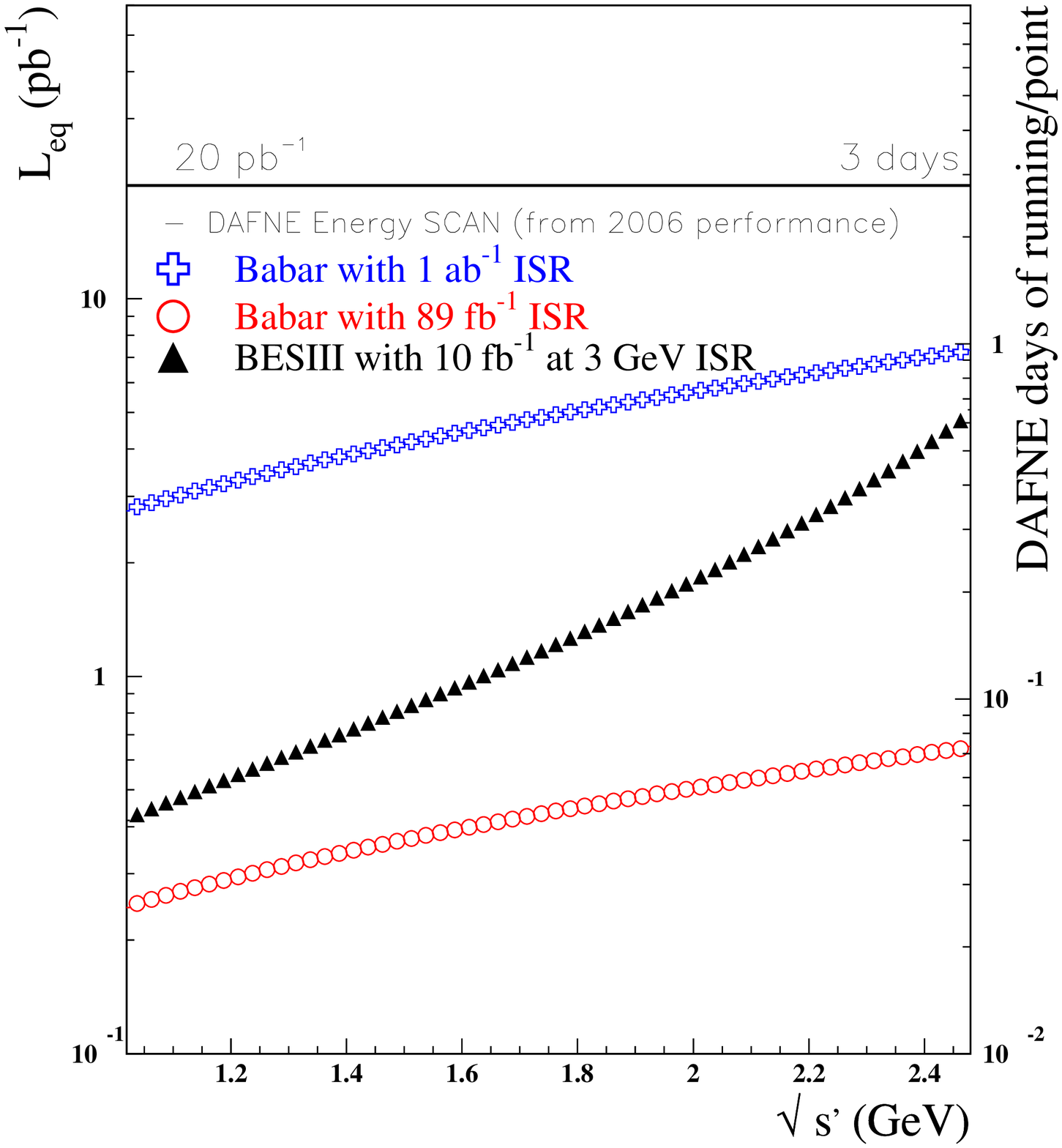}
\vspace{-1.5cm}
\figcaption{\label{fig:w2} Equivalent luminosity for: BaBar with 
1 ab$^{-1}$ 
(cross); BaBar with 89 fb$^{-1}$ (circle); 
BES-3 with 10 fb$^{-1}$, using ISR 
 at 3 GeV (triangle). A bin width of 25 MeV is assumed. 
A polar angle of the photon larger than $20^\circ$ is assumed.}
\end{center}

As shown in Fig.~\ref{fig:w2} the statistical yield 
 will be one order of magnitude higher 
than with 1 ab$^{-1}$ at BaBar, and better than BES-3.
%
%
Figure~\ref{fig:impactscan}  shows the statistical error for the channels
$\pi^+\pi^-\pi^0$, $2\pi^+ 2\pi^-$ and $\pi^+\pi^-K^+K^-$, which can be
 achieved by  an energy scan at DA$\mathrm{\Phi}$NE upgraded in energy
 with 20 pb$^{-1}$ per point,
compared  
with BaBar with published (89 fb$^{-1}$), and tenfold 
(890 fb$^{-1}$) statistics.
\begin{center}
\includegraphics[width=9cm,height=12cm]{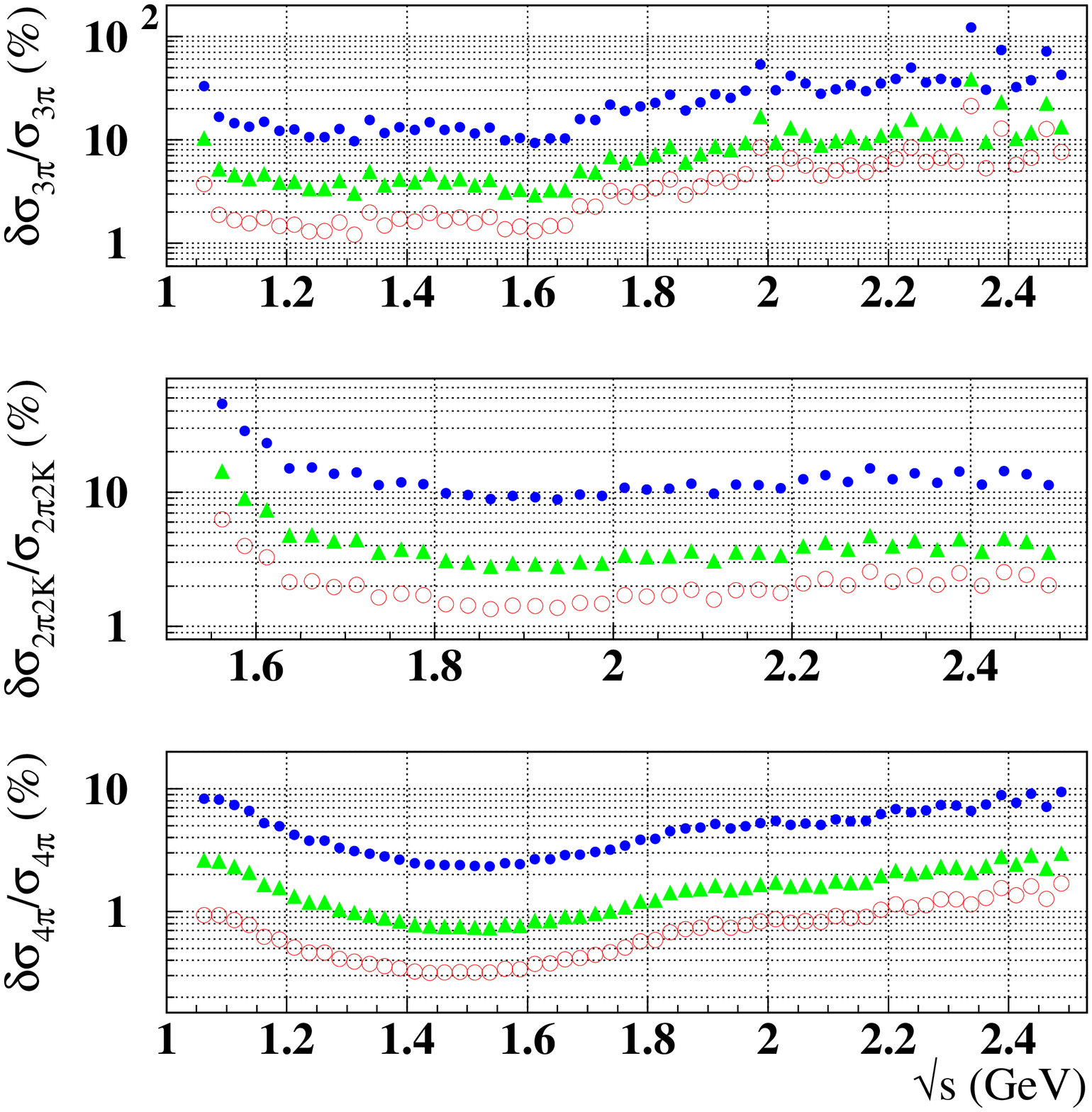}
\vspace{-1.cm}
\figcaption{\label{fig:impactscan} Comparison of the statistical accuracy in
the cross section among DA$\mathrm{\Phi}$NE upgraded in energy
 with an energy scan with 20 pb$^{-1}$ per 
point ($\circ$); published BaBar results ($\bullet$), 
BaBar with 890 pb$^{-1}$ statistics ($\blacktriangle$)  for $\pi^+\pi^-\pi^0$
(top), $\pi^+\pi^-K^+K^-$ (middle) and $2\pi^+ 2\pi^-$ (down) channels. 
An energy step of 25 MeV is assumed.}
\end{center}

As can be seen, an energy scan allows to reach a statistical
 accuracy of order of 1\% for most of the energy points.
(In addition, KLOE-2 can benefit from the high machine luminosity to
use ISR as well). 
The detector  hermeticity, the high granularity of the calorimeter, 
the excellent momentum resolution of the drift chamber and the insertion of the inner tracker would allow to reduce the systematic error to the same level. The beam energy can be determined with an error less than 100 keV,
by using  the Compton backscattering (CBS) of laser photons against the electron beam.

To summarize the impact for the $g-2$ of the muon, 
KLOE-2 by itself can bring the accuracy on $a^{\rm{HLO}}_{\mu}$ 
to about $2.5\times$10$^{-10}$, by measuring  the ratio  $\pi^+\pi^-(\gamma)$ to $\mu^+\mu^-(\gamma)$  with a 0.4\% accuracy at 1 GeV with ISR, and the hadronic cross sections in the region [1--2.5 GeV] with 1-2\% error.
 This would represent a factor two improvement on the current error of
$a^{\rm{HLO}}_{\mu}$, which is
 necessary in order to match the increased precision of the proposed muon g-2 experiments at FNAL~\cite{lee} and J-PARC~\cite{mibe}.

\subsection{Summary}
The status and the physics program of the KLOE-2 experiment has been discussed.
The detector has been upgraded with an electron tagger 
for $\gamma\gamma$ physics.
The insertion of an inner tracker and calorimeters in the forward regions are
 planned for the end of 2011.
KLOE-2 will start data taking in Spring 2010, 
and will take data for at least 3 years.
The physics program is wide, spanning from studies on neutral
kaon quantum interferometry to precise measurement 
of hadronic cross sections in the energy range below 2.5 GeV. 
It will have a major impact in many tests of the Standard Model (like the $g-2$ of the muon), tests of discrete symmetries, and searches for new physics.

\vspace{0.5cm}
\acknowledgments{I would like to thank D.~Babusci, C.~Bloise, F.~Bossi, A.~Di~Domenico, S.~Eidelman, P.~Gauzzi and F.~Nguyen for a careful reading of the manuscript and useful discussions.
Many thanks to the organizers of the workshop, especially Changzheng Yuan, for the warm hospitality and the pleasant and stimulating atmosphere of this conference.}

\end{multicols}
\clearpage
\end{document}